\documentclass[lettersize,journal]{IEEEtran}
\usepackage{ifpdf}
\usepackage{cite}
\usepackage{color}
\usepackage{amsthm}
\usepackage{amsmath,amssymb}
\usepackage{algorithmic}
\usepackage{array}
\usepackage{mdwmath}
\usepackage{mdwtab}
\usepackage{graphicx}
\usepackage{lipsum}
\usepackage{epstopdf }
\usepackage{algorithm,algorithmic}
\usepackage{enumerate}
\usepackage{eqparbox}
\usepackage{url}
\usepackage{subfig}
\usepackage{amsbsy}
\usepackage{multicol}

\usepackage{dblfloatfix}    
\usepackage{mathtools, cuted}

\usepackage{geometry}
\def\BibTeX{{\rm B\kern-.05em{\sc i\kern-.025em b}\kern-.08em T\kern-.1667em\lower.7ex\hbox{E}\kern-.125emX}}

\newcommand{\be}{\begin{equation}}
	\newcommand{\ee}{\end{equation}}
\newcommand{\beq}{\begin{eqnarray}}
	\newcommand{\eeq}{\end{eqnarray}}

\makeatletter
\newif\ifnobrackets
\renewcommand\@cite[2]{\ifnobrackets\else[\fi{#1\if@tempswa , #2\fi}\ifnobrackets\else]\fi\nobracketsfalse}

\makeatother

\newcommand\numeq[1]%
{\stackrel{\scriptscriptstyle(\mkern-1.5mu#1\mkern-1.5mu)}{=}}

\usepackage{amsmath,amsfonts}
\usepackage{algorithmic}
\usepackage{array}
\usepackage{lipsum}
\begin{document}

	\title{ \huge {5G} NR Positioning Enhancements in 3GPP Release-18}

	\author{\IEEEauthorblockN{Hyun-Su~Cha$^\dagger$, Gilsoo~Lee$^\dagger$, Amitava~Ghosh$^\dagger$, Matthew Baker$^\ddagger$, Sean Kelley$^\dagger$, and Juergen Hofmann$^{\dagger\dagger}$  
		} \\  
	\IEEEauthorblockA{{\small 
				$^\dagger$Nokia Standards, USA. Emails:\{hyun-su.cha, gilsoo.lee, amitava.ghosh, sean.kelley\}@nokia.com}  \\ 
		}
		\IEEEauthorblockA{ {\small
			$^\ddagger$Nokia Standards, UK. Email: matthew.baker@nokia.com }\\ 
		}
		\IEEEauthorblockA{ {\small
			$^{\dagger\dagger}$Nokia Standards, Germany. Email: juergen.hofmann@nokia.com }\\ 
		}
		
	}
	
	\maketitle
	\thispagestyle{empty}
	
\begin{table*}
	\boxed{\text{This work has been submitted to the IEEE for possible publication. Copyright may be transferred, after which this version may no longer be accessible }}
\end{table*}
	
\begin{abstract}	
New radio (NR) positioning in the Third Generation Partnership Project (3GPP) Release 18 (Rel-18) enables 5G-advanced networks to achieve ultra-high accuracy positioning without dependence on global navigation satellite systems (GNSS) with key enablers such as the carrier phase positioning technique, standardized for the first time in a cellular communications standard and setting a new baseline for future generations. In addition, Rel-18 NR supports positioning functionalities for reduced capability (RedCap) user equipment and bandwidth aggregation for positioning measurements. Moreover, the low power solutions are designed for low power high accuracy positioning use cases. Lastly, sidelink based positioning is introduced in Rel-18. This article constitutes a comprehensive treatment of the Rel-18 NR positioning enhancements crucial for the development of next-generation networks.
\end{abstract}

\vspace{-2mm}
\section{Introduction}

\begin{figure}[t]
	\centering
	\includegraphics[width=0.47\textwidth]{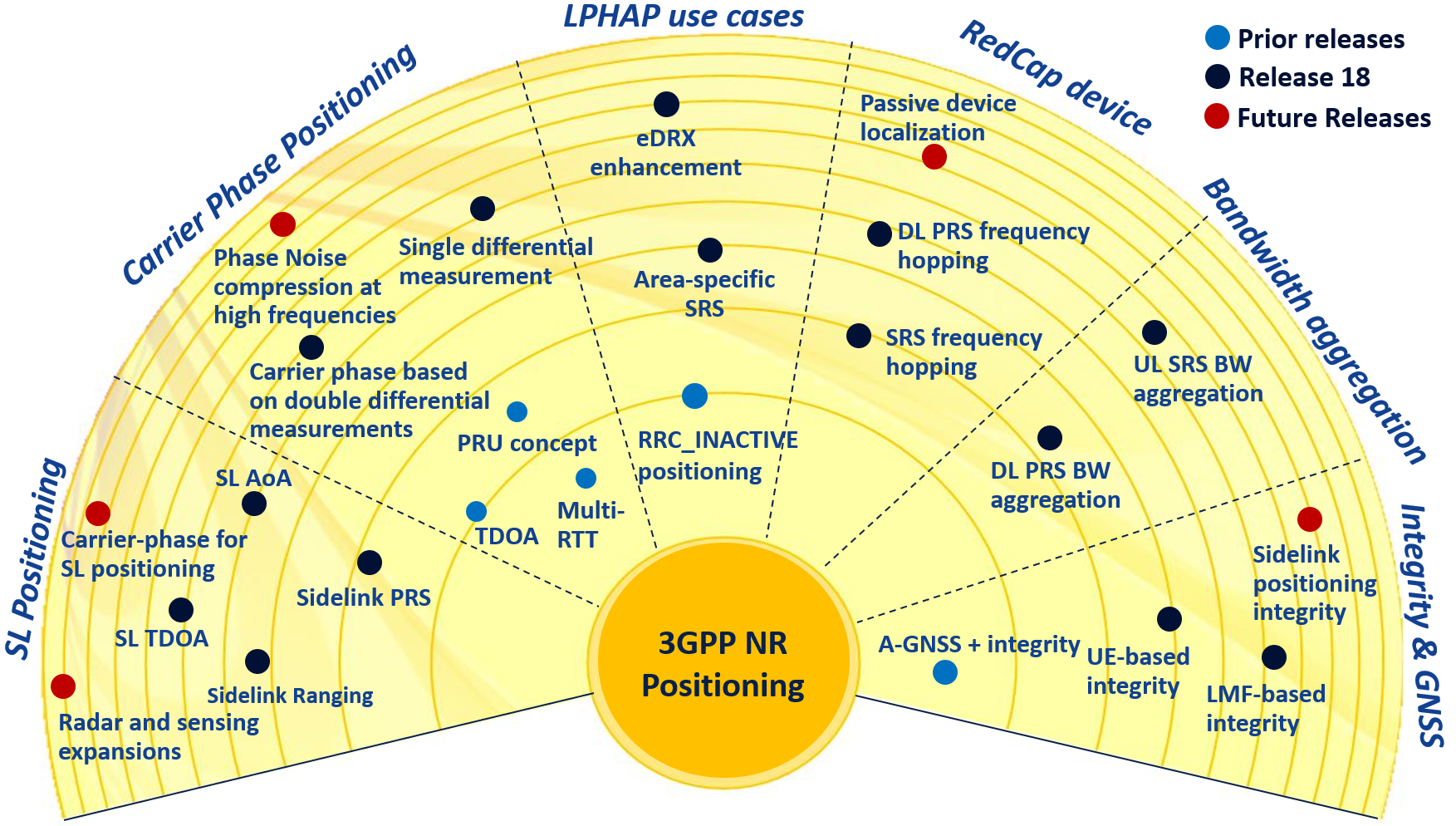} 
	\caption{{\color{black}Overview of 3GPP NR positioning} \vspace{-1mm}}
	\label{fig:overview}
\end{figure}

The Third Generation Partnership Project (3GPP) new radio (NR) positioning has come a long way since its inception, continuously evolving with innovative features and diverse measurement types. Initially in Release-15, positioning capability was provided by means of signaling to support techniques provided independently of 5G NR. Release 16 (Rel-16) laid the groundwork by introducing dowlink (DL) time difference of arrival (TDOA), DL angle of departure (AoD), uplink (UL) TDOA, UL angle of arrival (AoA), multiple round trip time (RTT), and enhanced cell identity techniques. DL positioning reference signals (PRS) and UL sounding reference signals (SRS) are tailored to location measurements of the introduced techniques. Fig.~\ref{fig:overview} shows an overview of the NR positioning.




As 5G networks advance and new industry verticals emerge, demanding applications require even more refined positioning performance. Release 17 (Rel-17) NR took on this challenge, focusing on accuracy, latency, network efficiency, and device efficiency for commercial, especially industrial internet of things (IIoT) use cases.

Release 18 (Rel-18) NR pushes the boundaries even further, unlocking the potential for applications in 5G-Advanced networks that demand ultra-high positioning accuracy – down to centimeter-level (cm-level) for IIoT use cases, and it lays a foundation to enable next-generation smart cities, encompassing enhanced emergency response, precise asset tracking, and immersive augmented reality experiences. Key features driving this leap include carrier phase positioning, reduced capability (RedCap) user equipment (UE) positioning support, and the introduction of sidelink (SL) positioning \cite{RP231460}.


The carrier phase positioning takes 5G-Advanced networks to a new level of location estimation accuracy, reaching the cm-level under certain radio channel conditions \cite{Fan2022}. The feasibility of cm-level positioning accuracy in NR using carrier phase is convincingly demonstrated by several sources including the technical report of Rel-18 NR positioning study \cite{38859}, and the work in \cite{Abdurrahman2022} which investigated high-precision positioning estimation for indoor factory deployment scenarios.



Rel-18 NR extends its reach to accommodate positioning functionality for RedCap UEs. These UEs have limited capability and complexity and operate with fewer radio frequency (RF) chains and strictly limited bandwidth (BW) of 20 MHz in frequency range 1 (FR1), i.e.,  sub-6 GHz frequency bands, and 100 MHz in FR2 \cite{ratasuk2021reduced}. To overcome this BW limitations, Rel-18 NR introduces frequency hopping for positioning measurements. This enables RedCap UEs to maintain satisfactory positioning accuracy, despite their BW limitations, opening up a wider range of application possibilities for these devices.



The accuracy of timing measurements such as reference signal time difference (RSTD) and Rx-Tx time difference is highly impacted by the BW size of RS. In FR1, the BW of DL PRS and UL SRS of normal NR UEs is limited to a maximum BW of 100 MHz. To overcome the limitation, Rel-18 NR introduces BW aggregation for positioning measurements. This feature allows Tx/Rx architectures with a single RF chain to receive and combine two or three intra-band contiguous component carriers (CCs), increasing the effective BW for positioning measurements. Through BW aggregation, the BW of DL PRS and UL SRS can be extended beyond the individual CC limit, leading to more accurate timing measurements.


The requirements for Low Power High Accuracy Positioning (LPHAP) are defined by the 3GPP Service and System Aspects Working Group 1 targeting IIoT scenarios such as massive asset tracking and automated guided vehicle tracking in industrial factories. 
A typical scenario is use case 6 of \cite{22104} that corresponds to tracking of workpieces in an assembly area or warehouse with a target accuracy of $<$1 m, a positioning interval of 15-30 seconds, and a battery life up to 1 year. For these industrial requirements, enhancements to extended discontinuous reception (eDRX) are introduced to help satisfy the battery life requirement, and the support of SRS valid in multiple cells facilitates power saving.



The 3GPP sidelink (SL) supports device-to-device communications for applications such as vehicle-to-everything (V2X) and emergency service communications when out of cellular coverage. Rel-18 expands upon the SL by introducing positioning functionality. SL PRS are designed to obtain SL positioning measurements and SL positioning protocol (SLPP) is introduced to exchange positioning-related information including measurements. SL positioning may be used to provide a positioning service for the out-of-coverage UEs, as well as increasing positioning accuracy by hybrid use together with UE-to-base-station positioning.



Section~\ref{sec:R18enh} of this paper explains in detail Rel-18 positioning enhancements between UEs and 5G base stations (gNB), and Section~\ref{sec:sl} explains SL positioning. The future research topics are discussed with conclusions in Section \ref{conclusion}.   \\


\vspace{-0mm}
\section{Rel-18 NR Positioning }\label{sec:R18enh}

This section provides the details of carrier phase positioning, bandwidth aggregation for positioning measurements, RedCap positioning and LPHAP use cases.


\subsection{Carrier Phase Positioning}\label{sec:cpp}

Carrier phase positioning harnesses a phase shift observed within an RS at a receiver. The distance between a transmitter and a receiver is expressed by $d=\lambda N +\phi_{DL}$, where $\lambda$, $N$, and $\phi_{DL}$, respectively, are the wavelength of a carrier frequency, the number of complete wavelength cycles traveled, and a DL reference signal carrier phase (RSCP) measurement in meters. The DL RSCP measurement, denoted by $\phi_{DL}=\hat\phi+\phi_r-\phi_t$, is subject to phase biases introduced by both the transmitter and receiver, where $\hat\phi$ is a phase value depending on the distance, $\phi_r$ is the receiving (Rx) phase bias from the receiver, and $\phi_t$ is the transmitting (Tx) phase bias from the transmitter. These phase bias needs to be canceled out by using differential measurements. 

\begin{figure*}[ht]
	\centering
	\includegraphics[width=0.80\textwidth]{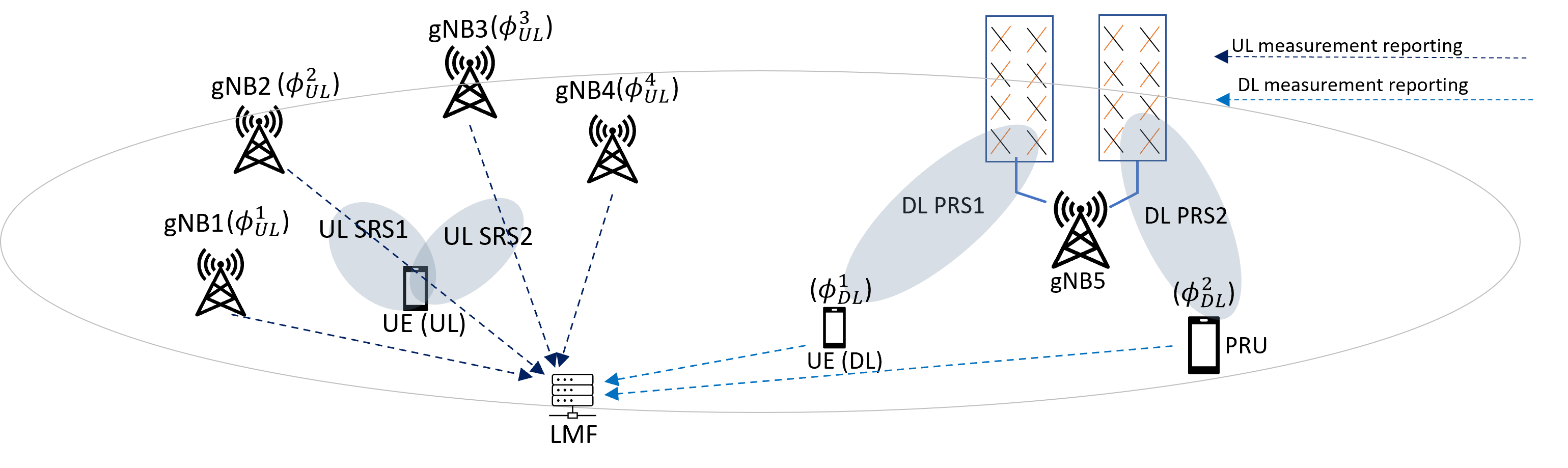} 
	\caption{{\color{black} Illustrative example of transmissions and receptions of DL PRS and UL SRS for carrier phase measurements.} \vspace{-1mm}}
	\label{fig:CP_DL_UL}
\end{figure*}

\textbf{DL:} Rel-18 NR positioning introduces a reference signal carrier phase difference (RSCPD) to exploit the difference of two RSCP measurements obtained from PRSs from two gNBs. The reporting granularity for DL RSCPD or DL RSCP is 0.1 degree, and the reporting range of DL RSCPD and DL RSCP is [-180, 180) and [0, 360) degrees, respectively. For example, 0.1 degree is corresponding to 0.24~mm if the carrier frequency is 3.5 GHz. UE-assisted positioning relies on RSCPD measurements reported by the UE to a location estimation entity, named location management function (LMF) \cite{37355}, effectively eliminating the Rx phase bias ($\phi_r$). To ensure identical Rx phase bias within these measurements, both gBNs must share the same DL PRS positioning frequency layer (PFL). This enables the UE to consistently employ the same Rx antenna, guaranteeing a consistent Rx phase bias for accurate subtraction.

Carrier phase positioning further necessitates a positioning reference unit (PRU) to remove the Tx phase bias. For example, a PRU may be a reference UE strategically positioned at a fixed, known location to the LMF. By ensuring identical Tx phase bias in DL PRS transmissions during the RSCPD measurement period for both the UE and the PRU, the LMF can successfully eliminate the Tx phase bias by calculating the difference of their respective RSCPD measurements.





Despite the use of a PRU, achieving identical Tx phase bias in RSCPD measurements for both target UE and PRU poses challenges. The factors contributing to discrepancies include the use of different antennas and measurement times \cite{nokia114_CP_POS}. As illustrated in Fig.~\ref{fig:CP_DL_UL}, if the PRU and target UE receive PRSs from different antenna panels, the resulting Tx phase biases for DL PRS 1 and DL PRS 2 transmitted at the same gNB can be different. Even for the same PRS, a phase drift can introduce varying phase biases if measurement timings between the PRU and target UE are not precisely aligned. To address this, the LMF can provide a measurement time window to align measurement timing and can indicate specific PRSs for the carrier phase measurements.



Rel-18 also supports UE-based carrier phase positioning where the UE-based positioning means the location estimation is done by the UE itself. By providing a UE with the PRU's RSCPD measurements, along with associated timestamps, the UE can compute a double differential measurement, effectively canceling out the Tx phase biases. Therefore, a UE can independently estimate its own location using the provided TRP locations and the measurements. The UE performs carrier phase measurements always in a measurement gap in RRC\_CONNECTED state and performs the measurements in time windows or selected DRX ON periods in RRC\_INACTIVE state.

\textbf{UL:} A gNB orchestrates UL RSCP measurements from one or more TRPs, laying the groundwork for cancellation of UE Tx phase bias. The LMF achieves this cancellation by calculating the difference between UL RSCP measurements from different TRPs, assuming a consistent Tx phase bias. However, guaranteeing this consistency presents challenges. For two UL SRSs transmitted from different antennas, RSCP measurements may contain different Tx phase biases. Even for a single antenna, phase bias can show temporal variations, necessitating careful synchronization of measurement timing. To avoid these, the LMF guides gNB measurement behavior. By providing a precise window for RSCP measurements, the LMF aims to minimize phase drift, ensuring a consistent Tx phase bias within the window. In addition, it can request the gNB to focus measurements on specific UL SRS resources, potentially reducing the impact of antenna-related bias inconsistencies.

Fig.~\ref{fig:CP_DL_UL} illustrates an example of UL carrier phase positioning. The transmission of UL SRS1 and UL SRS2 from different antennas introduces different Tx phase biases, denoted as $\phi_{t1}$ and $\phi_{t2}$, respectively. To address this, the LMF coordinates TRP measurements, by requesting gNB1 and gNB2 to perform the UL RSCP measurement from UL SRS1, and gNB3 and gNB4 may be requested to focus on UL RSCP measurement from UL SRS2. By calculating differential measurements within these gNB pairs, i.e., $\phi_{t1}$ versus $\phi_{t2}$, and $\phi_{UL}^3$ versus $\phi_{UL}^4$, the LMF successfully eliminates Tx phase bias. Without this careful operation, the Tx phase bias $\phi_{t1}$, $\phi_{t2}$ would inevitably degrade measurement accuracy.



\textbf{DL and UL}: Additionally, there is an approach called RTT-like carrier phase positioning. Unlike DL-only or UL-only methods, it bypasses the need for a dedicated PRU to remove the Tx phase bias by using both DL RSCP and UL RSCP measurements. Importantly, these measurements should possess the same Tx and Rx phase bias values but with opposite signs. To achieve the common phase bias, the UE and the gNB should each use a single local oscillator for their RS transmission and reception. This ensures both UL SRS transmission and DL PRS reception at the UE, as well as UL SRS reception and DL PRS transmission at the TRP, share the same phase bias values, enabling their cancellation. In such a case, a measured UL RSCP is expressed by $\phi_{UL}=\hat\phi-\phi_r+\phi_t $. The LMF can remove $\phi_r$ and $\phi_t$ from the sum of $\phi_{DL}$ and $\phi_{UL}$. However, the same local oscillator condition is not mandated by specifications.



\subsection{Bandwidth Aggregation for positioning}\label{sec:bwaggr}
BW Aggregation is designed to enhance accuracy by increasing the BW over which measurements are made. 



\textbf{DL:} The LMF informs a UE of the DL PRS resource sets to be combined so that the aggregated measurements are obtained from BW aggregation across intra-band contiguous multiple DL PRS positioning frequency layers, where each DL PRS resource set may comprise multiple DL PRS resources. Among the DL PRS resources within the DL PRS resource sets, there are \textit{linked} DL PRS resources so that the UE can obtain an aggregated measurement. More specifically, for the two or three DL PRS resource sets configured for BW aggregation, if DL PRS resources of different DL PRS resource sets satisfy some criteria such as a same QC defined in \cite{38214}, these DL PRS resources are \textit{linked} for BW aggregation. 
For the linked DL PRS resources, the UE assumes phase continuity between them, so the gNB should guarantee the phase continuity of the linked DL PRS resources. The gNB should use a single Tx chain and the same transmit antenna reference point for the PRS transmission of the linked DL PRS resources.

The LMF can indicate a reporting granularity. $k\in\{-6,-5,\cdots,-1\}$ is a set of the granularity of $2^{k}\times T_c$ introduced in Rel-18, where $T_c=0.5086$~ns. 
When the UE reports a measurement, it can either inform if the measurement is an aggregated measurement or report which DL PRS positioning frequency layers are used.

\textbf{UL:} For UL BW aggregation, a LMF may request a gNB to provide UL positioning measurements from aggregated SRSs across two or three UL CCs.

The gNB provides the UE with the configuration of SRS resource sets for the purpose of the BW aggregation. Similar to the DL case, among the UL SRS resources in the UL SRS resource sets, there are \textit{linked} UL SRS resources so that the UE understands which SRS resources would be used for bandwidth aggregation. More specifically, for the two or three UL SRS resource sets configured for bandwidth aggregation, if UL SRS resources of different UL SRS resource sets satisfy some criteria such as spatial relation information \cite{38214}, these UL SRS  resources are \textit{linked} for the BW aggregation for the UL positioning measurement. For the \textit{linked} UL SRS resources, the UE should ensure equal transmission power per subcarrier and phase continuity of the SRS transmission of the linked UL SRS resources across multiple component carriers, enabling the gNB to acquire aggregated RSTD, gNB Rx-Tx time difference, RSRP or RSRPP measurements.

If one of the linked SRSs is collided with a physical uplink shared channel (PUSCH) or physical uplink control channel (PUCCH) on a CC, the UE does not transmit all linked SRSs for positioning across the CCs. For improved network efficiency, the gNB can use a single DCI (Downlink Control Information) message or MAC-CE (Medium Access Control Element) to trigger or activate multiple aperiodic or semi-persistent SRS resource sets, respectively. 



\subsection{RedCap Positioning}\label{sec:redcap}

In Rel-18, features introduced in the previous releases of NR positioning are supported for RedCap positioning by means of frequency hopping. 

\begin{figure}[t]
	\centering
	\includegraphics[width=0.47\textwidth]{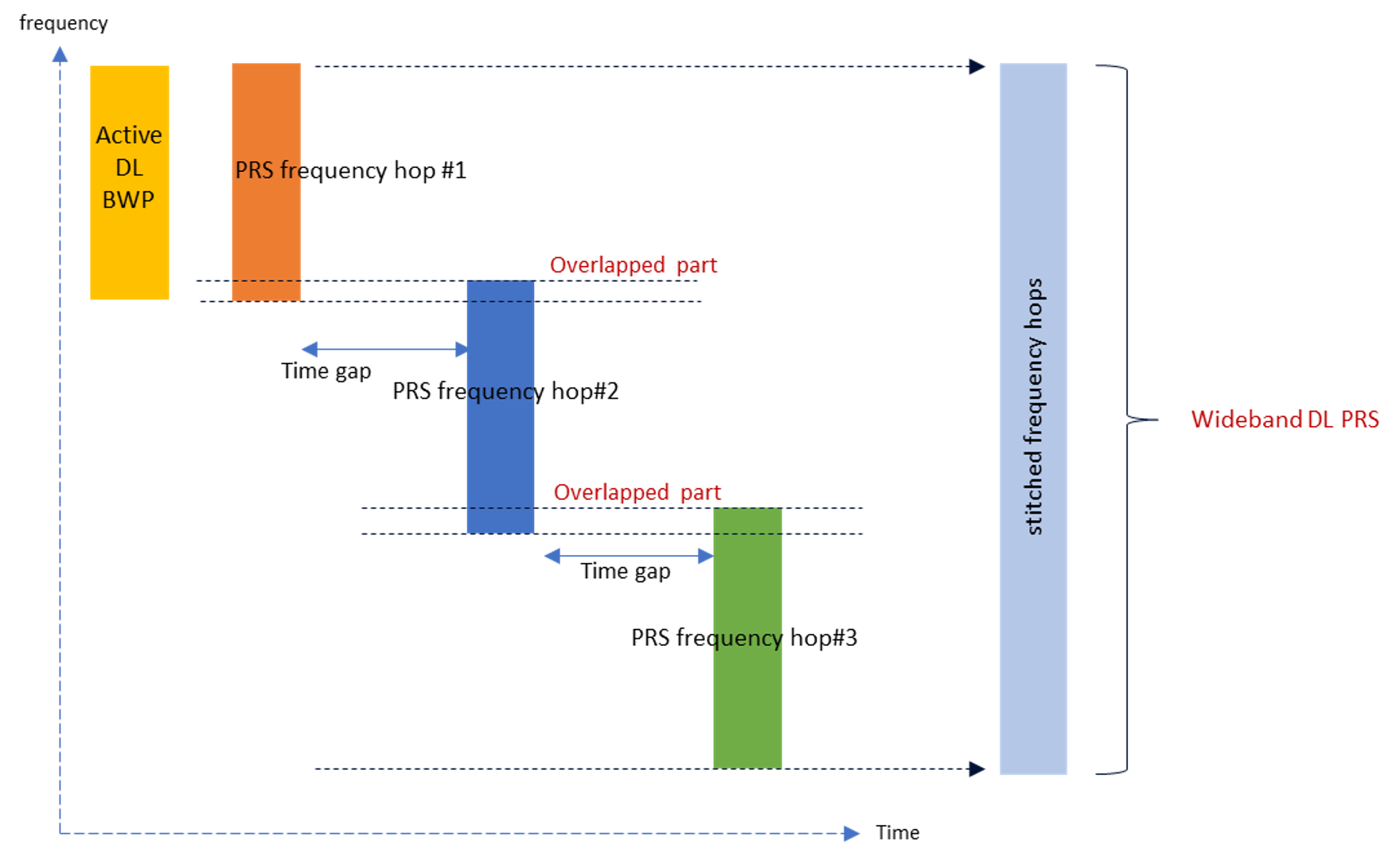} 
	\caption{{\color{black} Illustrative example of DL PRS frequency hopping} \vspace{-1mm}}
	\label{fig:PRS_FH}
\end{figure}

\begin{figure}[t]
	\centering
	\includegraphics[width=0.45\textwidth]{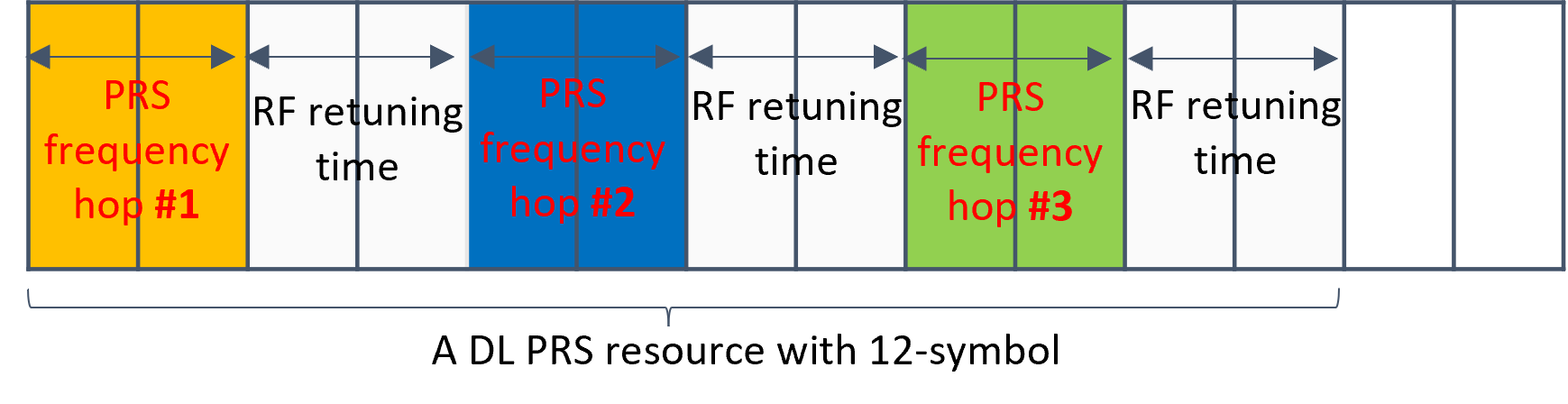} 
	\caption{{\color{black} Illustrative example timeline of the DL PRS frequency hopping with RF retuning.} \vspace{-1mm}}
	\label{fig:PRS_FH_time}
\end{figure}

\textbf{DL PRS frequency hopping:} RedCap UEs can receive a part of the wideband DL PRS, called as a frequency hop. By sequentially receiving multiple frequency hops and combining them, they can effectively process wideband DL PRS signals, despite their limited instantaneous maximum BW.
However, there are challenges such as RF retuning and phase differences. RedCap UEs must retune their RF to receive different frequency hop, resulting in phase differences between the frequency hops. To accurately reconstruct the wideband DL PRS, these phase differences must be compensated. Fig.~\ref{fig:PRS_FH} illustrates one solution introduced in Rel-18. By receiving DL PRS from overlapping resource blocks in adjacent frequency hops, UEs can estimate and correct for the phase difference. Fig.~\ref{fig:PRS_FH_time} illustrates the timeline of the frequency hopping within a 12-symbol DL PRS resource. The UE receives the DL PRS from a frequency hop over two symbols and performs RF re-tuning in the following two symbols. For a gNB that is sufficiently close to the UE, two symbols per frequency hop can be enough. However, longer reception periods are likely to be needed for each frequency hop to ensure adequate signal strength if gNBs are further away, potentially resulting in longer delays.



RedCap UEs have flexibility to operate either with or without frequency hopping. In particular, UEs can acquire positioning measurements from a single frequency hop of the DL PRS, bypassing the complexity of RF retuning and frequency hopping. On the other hand, they can leverage frequency hopping to construct a wider bandwidth DL PRS, although this needs measurement gaps to be configured in the data reception bandwidth in order to retune to make the measurements on other frequencies. In RRC\_CONNECTED state, DL PRS frequency hopping are always done within a measurement gap, whilst in RRC\_INACTIVE state, the UE performs it in selected DRX ON periods.

 UEs may inform the LMF whether measurements are from a single or multiple frequency hops. This enables the LMF to prioritize measurements derived from multiple hops for better accuracy, if desired.




\textbf{UL SRS frequency hopping:} RedCap UEs can use UL SRS frequency hopping to transmit wideband UL SRS signals in segments, called frequency hops, such as 20 MHz out of 100 MHz SRS in FR1. SRS frequency hopping necessitates a separate SRS configuration including virtual frequency resource allocation, BW per frequency hop, starting PRB, overlapping RBs, periodicity, offsets, and time duration of each frequency hop. gNBs can provide time windows with a length of 1, 2, 4, or 6 slots to prioritize SRS frequency hopping over data communications. This ensures uninterrupted SRS transmission in all frequency hops, which is crucial for gNBs to accurately reconstruct the wideband SRS. However, if the window is not provided, the other UL signals are mostly prioritized over SRS transmission, which facilitates stable data communications but could result in a delay to obtain positioning measurements. 


\begin{figure}[t]
	\centering
	\includegraphics[width=0.43\textwidth]{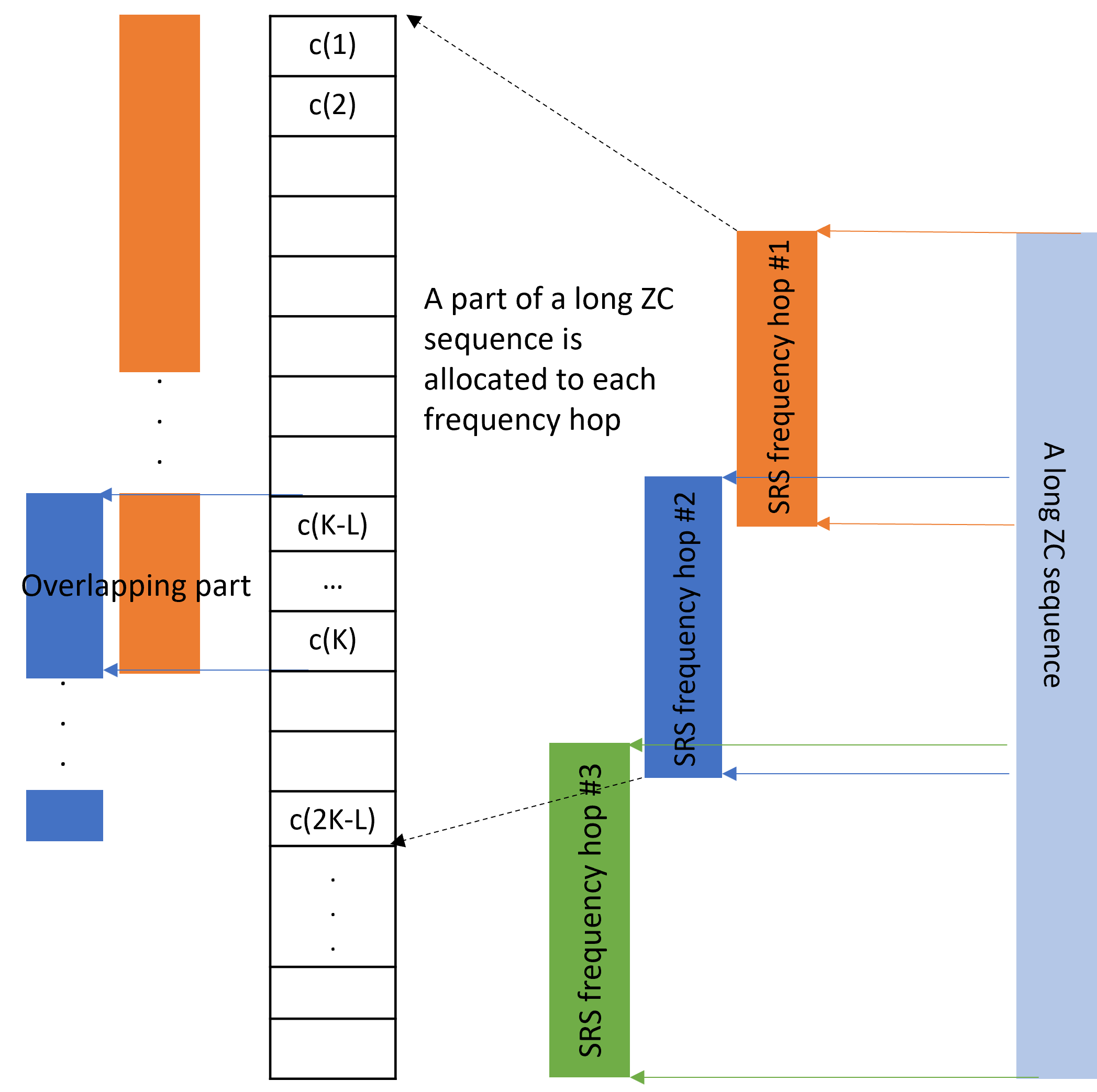} 
	\caption{{\color{black} Illustrative example of UL SRS frequency hopping with Zadoff-Chu sequence mapping.} \vspace{-1mm}}
	\label{fig:SRS_FH}
\end{figure}

A single ZC sequence for the SRS transmission is generated based on subcarriers of the overall SRS frequency hops, even though the RedCap UE cannot transmit the whole ZC sequence at the same time. That is, although there are frequency resources that the UE cannot use at a specific time, the frequency resources are virtually allocated for the SRS frequency hopping \cite{nokia114_RedCap_POS}. Adjacent frequency hops intentionally share a portion of the ZC sequence elements, which is illustrated in Fig.~\ref{fig:SRS_FH} with a length $L$, enabling gNBs to estimate and compensate for phase differences between the frequency hops. Similar to DL PRS frequency hopping, to optimize measurement utilization, gNBs explicitly indicate to LMF whether measurements were derived from single or multiple frequency hops.


\subsection{LPHAP Use Case}\label{sec:lphap}
The area-specific SRS configuration valid in multiple cells is a new feature for UEs in RRC\_INACTIVE state to save power. By adopting an area-specific SRS configuration, a UE can continue using a given configured SRS for positioning when it moves to another cell within the validity area. Since the validity area can include cells of multiple gNBs, the LMF may coordinate the time and frequency resources for SRS within the validity area to avoid interference. Within the validity area, the UE can autonomously adjust the timing advance when cell-reselection occurs. In that case, it is required for a UE to maintain the same UL SRS transmission timing to mitigate UL RTOA measurement errors.
In addition, new positioning measurement requirements are adopted for eDRX cycles beyond 10.24 sec for LPHAP so that the UE can save power by staying in sleep mode. When a UE is configured to measure PRS or transmit SRS for positioning, it can perform RRM measurements for positioning outside PTW if eDRX cycle is larger than the configured measurement reporting periodicity or SRS transmission periodicity. In addition, if eDRX cycles and DL PRS measurement occasions are not aligned in time domain, a UE in RRC\_INACTIVE state may not be able to sleep or miss the chance to perform positioning measurements. Thus, NR supports  timeline alignment of eDRX cycles and DL PRS measurement.

\section{Rel-18 NR Sidelink Positioning}\label{sec:sl}

Rel-18 NR positioning introduced SL PRS to enable UE to measure and report positioning measurements. It supports SL TDOA, SL RTOA, SL RTT, and SL AOA techniques. 

\textbf{SL PRS:} SL PRS is based on Gold sequence, and the supported subcarrier spacing (SCS) for SL PRS is 15 kHz, 30 kHz, 60 kHz for FR1, and 60 or 120 kHz for FR2. 

A SL PRS resource refers to a time and frequency resource within a slot used for the transmission and reception of the SL PRS. The SL PRS resource configuration comprises a frequency-domain comb pattern of length 1, 2, 4, or 6, across 1 to 8 symbols \cite{38211}. The resource element (RE) patterns of the SL PRS are fully staggered enable the UE to combine the received SL PRS across multiple symbols into one symbol contiguous subcarriers (i.e., a comb pattern length of 1). There are two types of resource pool for SL PRS configurations. In a shared resource pool, the resource can be shared between SL data communications and the SL positioning, while a dedicated resource pool is only for SL positioning. A Physical Sidelink Control Channel (PSCCH) in the dedicated resource pool carries Sidelink Control Information (SCI) associated with the SL PRS resources. 

It is expected to support in-coverage, out-of-coverage, and partial coverage scenario. The latter being a scenario where a UE is out of base-station coverage but within SL coverage of a UE that is within base-station coverage.
Requirements for measurements from SL PRS are specified up to 40MHz BW for FR1. The SL positioning measurement period requirements such as SL RSTD, SL AoA, and SL RTOA are defined as number of slots of SL PRS resource(s) and minimum measurement processing time.  
A single measurement sample is allowed for SL PRS with a BW of more than 48 physical resource blocks (PRBs) while four samples are required if SL PRS BW ranges between 24 and 48 PRBs. 



\begin{figure}[t]
	\centering
	\includegraphics[width=0.45\textwidth]{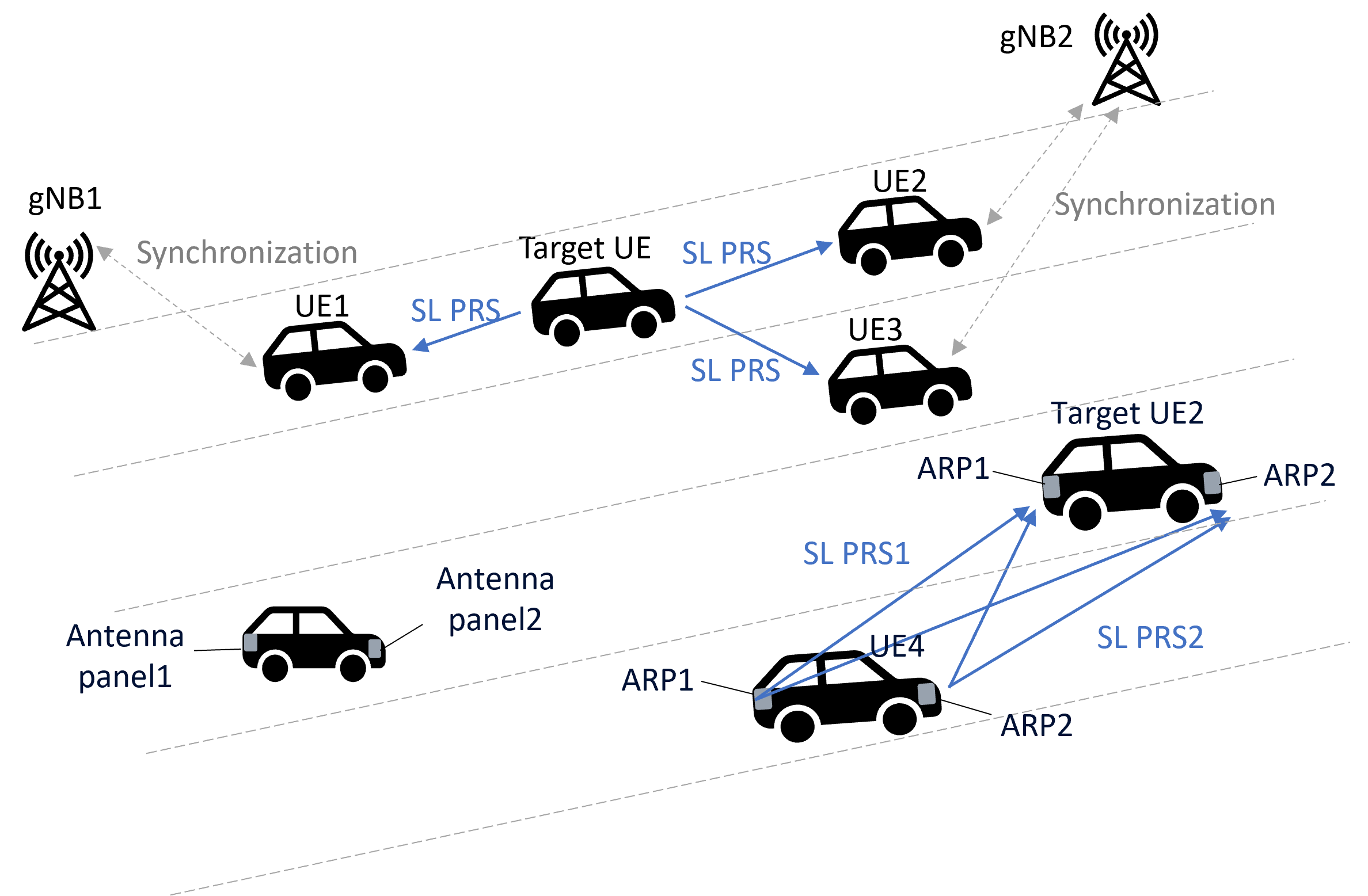} 
	\caption{{\color{black} UE1 is synchronized to gNB1, and UE2 and UE3 are synchronized to gNB2. The UEs are equipped with two antenna panels mapped to ARP1 and ARP2 respectively.} \vspace{-1mm}}
	\label{fig:SL_POS}
\end{figure}

\textbf{SL-TDOA:} 
In DL-like SL-TDOA technique, a target UE performs SL RSTD measurements, which comprise the propagation time difference of two different SL PRSs transmitted from different UEs. One of the SL PRSs transmitted from the UEs would be a reference of the SL RSTD measurements, which is reported by the target UE. Only the anchor UEs need to transmit SL PRSs. In contrast, UL-like SL-TDOA technique requires SL PRS transmission from a target UE and demands anchor UEs to measure SL RTOA measurements. However, UL-like SL-TDOA technique requires only a single SL PRS transmission at the target UE. Thus, it provides higher SL PRS resource efficiency than for DL-like SL-TDOA. In SL communications, SL UEs can use different synchronization sources as shown in Fig.~\ref{fig:SL_POS}, where gNB 1 is a synchronization source of UE~1 while gNB~2 is a synchronization source to UE~2 and UE~3. To compensate for time errors from different synchronization sources, SL positioning measurements reported by UEs may be accompanied by the used synchronization source information \cite{nokia113_measurement_SL_POS}.

\textbf{SL-RTT:} For the SL-RTT positioning technique, both the target UE and anchor UEs need to measure and report UE Rx-Tx time difference measurements so that the LMF can obtain a round trip time. For improved measurement accuracy, the UE can use an actual SL PRS transmission timing to obtain a UE Rx-Tx measurement if the reference SL PRS transmission timing changes while performing the UE Rx-Tx time difference measurement. In addition, for the mitigation of the UE Rx-Tx measurement error due to timing drift, SL RTT supports multiple UE Rx-Tx time difference measurements for a single SL PRS reception or a single SL PRS transmission, known as double-sided SL-RTT.
SL-RTT is robust to time synchronization errors, but it inherently comes with more delay and low RS efficiency than SL-TDOA.


\textbf{SL-AoA:} 
A horizontal angle and/or vertical angle measurements from an SL PRS are used for SL-AoA. Either a global coordinate system (GCS) or a local coordinate system (LCS) can be used. The translation of the LCS to the GCS can be done by a LMF. With an LCS, the estimated azimuth angle is measured relative to the x-axis of the LCS and is positive in a counter-clockwise direction, and an estimated vertical angle is measured relative to the z-axis of LCS and is positive towards the x-y plane direction. The bearing, downtilt and slant angles of LCS are defined in \cite{38901}. 


\textbf{Multiple antenna panels:} Vehicular UEs typically has enough space to have multiple antennas, as illustrated in Fig.~\ref{fig:SL_POS}. The UEs may be able to perform a SL positioning measurement per antenna panel. The measurements from different antenna panels would be different due to spatial separation between antennas \cite{nokia112bis_measurement_SL_POS}. For example, an antenna panel may not see a line of sight path to another UE, while another one may see it. To support measurement reporting per antenna panel, antenna reference point (ARP) concept is defined and associated with SL positioning measurements. For instance, Fig.~\ref{fig:SL_POS} shows vehicle UEs equipped with two antenna panels in the front and the rear, respectively that are mapped to ARP1 and ARP2. The positioning measurements may be different depending on ARP, e.g., horizontal angles measured from ARP1 and ARP2 of target UE2 may be different.
Therefore, target UE2 needs to report the ARP identity associated with SL PRS measurements and/or a transmitted SL PRS to avoid the location estimation accuracy degradation. The ARP-specific measurement reporting enables the LMF to prioritize measurements associated with an specific ARP and to estimate ARP-specific location of a UE. 





\vspace{-1mm}
\section{Conclusion and Future Work}\label{conclusion}

The NR positioning enhancements of Rel-18 hold immense promise for the future of high-precision location services. In this article, we have provided the first comprehensive summary on these advancements, including carrier phase positioning that is standardized for the first time in a cellular communication system to provide cm-level accuracy in 5G-Advanced networks. The new positioning features also include bandwidth aggregation for positioning measurements, the adoption of frequency hopping for RedCap UEs, low-power consumption solutions for LPHAP use cases, and the sidelink positioning. As a continued effort of the Rel-18 work, future 3GPP releases will need innovative enhancements. 

These technologies will now be the baseline against which future technologies and generations will be compared. Further developments of these techniques might address the fact that carrier phase positioning, bandwidth aggregation, and DL PRS frequency hopping are currently applicable only within measurement gap configurations in RRC\_CONNECTED state. For the carrier phase positioning, the impact of timing error group defined in \cite{38214} to the measurements needs to be considered.
 For RedCap positioning with SRS frequency hopping, the PAPR performance needs to be investigated, as the UE transmits a part of ZC sequence at a hop, which may impact the coverage of SRS frequency hopping. Moreover, various power reduction strategies such as skipping measurement or reporting identified in the study item phase \cite{38859} were not included in the scope of Rel-18. Further optimization of SL positioning may be needed in the future for frequencies higher than FR1, and techniques to support SL PRS in unlicensed bands is also deemed as future work.

\vspace{-2mm}
\bibliography{Pos.bib}{}

\begin{thebibliography}{10}
\providecommand{\url}[1]{#1}
\csname url@samestyle\endcsname
\providecommand{\newblock}{\relax}
\providecommand{\bibinfo}[2]{#2}
\providecommand{\BIBentrySTDinterwordspacing}{\spaceskip=0pt\relax}
\providecommand{\BIBentryALTinterwordstretchfactor}{4}
\providecommand{\BIBentryALTinterwordspacing}{\spaceskip=\fontdimen2\font plus
\BIBentryALTinterwordstretchfactor\fontdimen3\font minus
  \fontdimen4\font\relax}
\providecommand{\BIBforeignlanguage}[2]{{%
\expandafter\ifx\csname l@#1\endcsname\relax
\typeout{** WARNING: IEEEtran.bst: No hyphenation pattern has been}%
\typeout{** loaded for the language `#1'. Using the pattern for}%
\typeout{** the default language instead.}%
\else
\language=\csname l@#1\endcsname
\fi
#2}}
\providecommand{\BIBdecl}{\relax}
\BIBdecl

\bibitem{RP231460}
{{RP-}233382}, ``Revised {WID} on {E}xpanded and {I}mproved {NR}
  {P}ositioning,'' Jun. 2023, 3GPP TSG-RAN Meeting \#100.

\bibitem{Fan2022}
S.~Fan, W.~Ni, H.~Tian, Z.~Huang, and R.~Zeng, ``Carrier phase-based
  synchronization and high-accuracy positioning in {5G} new radio cellular
  networks,'' \emph{{IEEE} Transactions on Communications}, vol.~70, no.~1, pp.
  564--577, Oct. 2022.

\bibitem{38859}
{{TR} 38.859}, ``{S}tudy on expanded and improved {NR} {P}ositioning
  {E}nhancements ({R}elease 18),'' Jan. 2023, v18.0.0.

\bibitem{Abdurrahman2022}
A.~Fouda, R.~Keating, and H.-S. Cha, ``Toward cm-{L}evel accuracy: Carrier
  phase positioning for {IIoT} in {5G-A}dvanced {NR} networks,'' in \emph{Proc.
  IEEE International Symposium on Personal, Indoor and Mobile Radio
  Communications (PIMRC)}, 2022, pp. 782--787.

\bibitem{ratasuk2021reduced}
R.~Ratasuk, N.~Mangalvedhe, G.~Lee, and D.~Bhatoolaul, ``{R}educed {C}apability
  {D}evices for {5G} {IoT},'' in \emph{Proc. {IEEE} PIMRC}, 2021, pp.
  1339--1344.

\bibitem{22104}
{{TS} 22.104}, ``{S}ervice requirements for cyber-physical control applications
  in vertical domains; {S}tage 1,'' Dec. 2021, v18.3.0.

\bibitem{37355}
{{TS} 37.355}, ``{LTE} {P}ositioning {P}rotocol ({LPP}),'' Jan. 2024, v17.7.0.

\bibitem{nokia114_CP_POS}
{R1}-2306819, ``{V}iews on {NR} {DL} and {UL} carrier phase positioning,'' 3GPP
  TSG-RAN WG1 Meeting \#114, Aug. 2023.

\bibitem{38214}
{3GPP {TS}38.214}, ``{NR};{P}hysical layer procedures for data ({R}elease
  18).''

\bibitem{nokia114_RedCap_POS}
{R1}-2306822, ``{V}iews on {P}ositioning for {R}ed{C}ap {UE}s,'' 3GPP TSG-RAN
  WG1 Meeting \#114, Aug. 2023.

\bibitem{38211}
{3GPP {TS}38.211}, ``{NR};{P}hysical channels and modulation ({R}elease 18).''

\bibitem{nokia113_measurement_SL_POS}
{R1}-2304346, ``{O}n measurements and reporting for {SL} positioning,'' 3GPP
  TSG-RAN WG1 Meeting {\#}113, May 2023.

\bibitem{38901}
{{TR} 38.901}, ``{S}tudy on channel model for frequencies from 0.5 to 100
  {GH}z,'' Mar. 2022, v17.0.0.

\bibitem{nokia112bis_measurement_SL_POS}
{Nokia, Nokia Shanghai Bell}, ``{O}n measurements and reporting for {SL}
  positioning,'' 3GPP TSG-RAN WG1 Meeting {\#}113, {R1}-2302294, April 2023.

\end{thebibliography}
\bibliographystyle{IEEEtran}

\end{document}